\documentclass[twocolumn,showpacs,aps,prl,superscriptaddress]{revtex4}

\usepackage{graphicx}
\usepackage{dcolumn}
\usepackage{amsmath}
\usepackage{amssymb}
\usepackage{amsgen}
\usepackage{amsfonts}
\usepackage{epsfig}
\usepackage{xspace}
\usepackage{subfigure}

\usepackage{epstopdf} 

\input babarsym

\def\er #1 #2 { $#1 \pm #2$ }

\def\btn {\ensuremath {\B^+ \to \tau^+ \nu_{\tau}}\xspace}

\def\tautoenunu {\ensuremath {\tau^+ \to e^+ \nu_e \overline{\nu}_\tau}\xspace}
\def\tautoe {\ensuremath {\tau^+ \to e^+ \nu_e \overline{\nu}_\tau}\xspace}

\def\tautomununu {\ensuremath {\tau^+ \to \mu^+ \nu_{\mu} \overline{\nu}_{\tau}}\xspace}
\def\tautomu {\ensuremath {\tau^+ \to \mu^+ \nu_{\mu} \overline{\nu}_{\tau}}\xspace}

\def\tautopinu {\ensuremath {\tau^+ \to \pi^+ \overline{\nu}_{\tau}}\xspace}
\def\tautopi {\ensuremath {\tau^+ \to \pi^+ \overline{\nu}_{\tau}}\xspace}

\def\tautopipiznu {\ensuremath {\tau^+ \to \rho^{+} \overline{\nu}_{\tau}}\xspace}

\def\tautorho {\ensuremath {\tau^+ \to \pi^+ \pi^{0} \overline{\nu}_{\tau}}\xspace}

\def\btodlnu      {\ensuremath{B^{-}\rightarrow\Dz\ell^{-}\bar{\nu} X}\xspace}

\def\eextra       {\ensuremath{E_{\mbox{\scriptsize{extra}}}}}

\def\NBG          {\ensuremath{N_{\mathrm{bg}}}}
\def\Nobs          {\ensuremath{N_{\mathrm{obs}}}}
\def\Nsig         {\ensuremath{N_{\mathrm{sig}}}}

\def\pprimesignal {\ensuremath{p^{\prime}_{{\mathrm{sig}}}}}

\def\bln {\ensuremath {\B^+ \to \ell^+ \nu_{\ell}}\xspace}
\def\ben {\ensuremath {\B^+ \to e^+ \nu_{e}}\xspace}
\def\bmun{\ensuremath {\B^+ \to \mu^+ \nu_{\mu}}\xspace}

\def\lhrbb  {\ensuremath {LHR_{B \overline{B}}}}
\def\lhrcont{\ensuremath {LHR_{\mathrm{cont.}}}}

\def\nBB    {\ensuremath {(458.9 \pm 5.1) \times 10^{6}} }

\def\onlumi {\ensuremath {417.6  \invfb\ }}

\def\btnsig {\ensuremath{2.3 \sigma}}
\def\btnpower{\ensuremath{\times 10^{-4}}}
\def\btntautoerawresult {\ensuremath {\left(3.6 \pm 1.4  \right)} }
\def\btntautomurawresult {\ensuremath {\left(1.3^{+1.8}_{-1.6}  \right)} }
\def\btntautopirawresult {\ensuremath {\left(0.6^{+1.4}_{-1.2}  \right)} }
\def\btntautorhorawresult {\ensuremath {\left(2.1^{+2.0}_{-1.8}  \right)} }

\def\btnrawresult {\ensuremath {\left( 1.7 \pm 0.8 \pm 0.2  \right)} }
\def\btnresult {\ensuremath {\btnrawresult\btnpower} }

\def\btnresultrawcomb {\ensuremath {\left(1.7 \pm 0.6  \right)} }
\def\btnresultcomb {\ensuremath {\btnresultrawcomb\btnpower} }
\def\combsig{\ensuremath {2.8 \sigma}}

\def\benlimit  {\ensuremath {0.8 \times 10^{-5}} }
\def\bmunlimit  {\ensuremath {1.1 \times 10^{-5}} }

\def\fBSq {\ensuremath{f_{B}^{2}}}
\def\fBSqresult {\ensuremath {(62 \pm 31) \times 10^{3} \mev^2} }
\def\Vub   {\ensuremath  (3.94 \pm 0.26) \times 10^{-3}}

\newcommand{\BABARPubYear}    {09}
\newcommand{\BABARPubNumber}  {029}

\newcommand{\SLACPubNumber} {13866}

\newcommand{\twowidefig} {0.48}

\def\figurebox#1#2#3{%
    \def\arg{#3}%
    \ifx\arg\empty
    {\hfill\vbox{\hsize#2\hrule\hbox to #2{\vrule\hfill\vbox to #1{\hsize#2\vfill}\vrule}\hrule}\hfill}%
    \else
    {\hfill\epsfbox{#3}\hfill}%
    \fi}

\begin{document}

\preprint{\babar-PUB-\BABARPubYear/\BABARPubNumber} 
\preprint{SLAC-PUB-\SLACPubNumber} 

\begin{flushleft}
\babar-PUB-\BABARPubYear/\BABARPubNumber\\
SLAC-PUB-\SLACPubNumber\\[10mm]
\end{flushleft}

\title{
{\large \bf
A Search for {\boldmath $\bln$} \xspace Recoiling Against {\boldmath \btodlnu}} 
}

%
\author{B.~Aubert}
\author{Y.~Karyotakis}
\author{J.~P.~Lees}
\author{V.~Poireau}
\author{E.~Prencipe}
\author{X.~Prudent}
\author{V.~Tisserand}
\affiliation{Laboratoire d'Annecy-le-Vieux de Physique des Particules (LAPP), Universit\'e de Savoie, CNRS/IN2P3,  F-74941 Annecy-Le-Vieux, France}
\author{J.~Garra~Tico}
\author{E.~Grauges}
\affiliation{Universitat de Barcelona, Facultat de Fisica, Departament ECM, E-08028 Barcelona, Spain }
\author{M.~Martinelli$^{ab}$}
\author{A.~Palano$^{ab}$ }
\author{M.~Pappagallo$^{ab}$ }
\affiliation{INFN Sezione di Bari$^{a}$; Dipartimento di Fisica, Universit\`a di Bari$^{b}$, I-70126 Bari, Italy }
% ...
\collaboration{The \babar\ Collaboration}
\noaffiliation

\date{\today}

\begin{abstract}
We present a search for the decay $\bln\ (\ell = \tau, \mu, \mathrm{or}\ e)$ in \nBB \xspace $\BB$ pairs
recorded with the \babar\ detector at the PEP-II $B$-Factory. 
We search for these \B\ decays in a sample of $\Bu\Bub$ events where one
\B-meson is reconstructed as \btodlnu. Using the method of Feldman and
Cousins, we obtain $\mathcal{B}(\btn) = \btnresult$, 
which excludes zero at \btnsig. We interpret the central value
in the context of the Standard Model and find the \B\ meson decay
constant to be $\fBSq = \fBSqresult$. We find no evidence for \ben\
and \bmun\ and set upper limits at the 90\% C.L. 
$\mathcal{B}(\ben)  < \benlimit$ and $\mathcal{B}(\bmun)  < \bmunlimit$.
\end{abstract}

\pacs{13.20.-v, 13.25.Hw, 12.15.Ji}

\maketitle

In the Standard Model (SM), the purely leptonic decay 
\bln~\cite{ChargeConjugation}
proceeds via quark annihilation into a $W^{+}$ boson. This 
process is related to the Cabibbo-Kobayashi-Maskawa matrix
element $V_{ub}$ and the \B\ meson decay constant, $f_{B}$,  
by $\BR(\bln) \propto |V_{ub}|^2f_{B}^2$. It is also potentially
sensitive to the presence of a charged Higgs boson~\cite{HiggsPaper}, as in the
minimal supersymmetric extension of the Standard Model. 
Using  $|V_{ub}| = \Vub$~\cite{HFAG} and 
$f_{B}~=~190\pm 13$~MeV~\cite{Gamiz:2009ku} and 
assuming only a SM contribution to the process, the
branching fraction predictions are
$\BR(\btn) = (1.0 \pm 0.2)\times 10^{-4}$, 
$\BR(\bmun) = (4.5 \pm 1.0)\times 10^{-7}$, and 
$\BR(\ben) = (1.1 \pm 0.2)\times 10^{-11}$.  
The different branching fractions result
from helicity suppression of the lower-mass charged 
leptons. In this paper, we describe a search
for all three final states.

The data used in this analysis were collected with the \babar\ detector
at the \pep2\ storage ring at the SLAC National Accelerator Laboratory. 
We use the full \babar\ dataset, corresponding to an integrated
luminosity of \onlumi\ with center-of-mass (CM) 
energy equal to the  \FourS\ rest mass.
These data contain
$\nBB$  $\FourS \to \BB$ pairs and we assume equal production of
\BzBzb\ and $\Bp\Bm$ from the \FourS\ decays.  
The \babar\ detector is described in detail elsewhere~\cite{babar}. 
For the most recent 203 \invfb\ of data, the barrel region of the 
muon system was upgraded to limited streamer 
tubes~\cite{Menges:2006xk}.

Signal and background processes
are simulated using EVTGEN~\cite{evtgen}. 
A GEANT4-based \cite{geant4} Monte Carlo (MC) 
simulation is used to model the detector response
and to estimate the signal efficiency and the physics backgrounds. 
Simulation samples
equivalent to approximately three times the accumulated data  were
used to model \BB\ events, and samples equivalent to approximately
1.5 times the accumulated data were used to model continuum background events
where $\epem \to$ \uubar, \ddbar, \ssbar, \ccbar, and \tautau. 
We independently simulate the signal processes at a rate 
over a hundred times that expected in data, using 
samples where one \B\ meson always decays as \bln\ 
and the second decays into any final state. We normalize these signal
samples to their predicted SM branching fractions.

The strategy adopted for this analysis is similar to that from our previously
published work~\cite{Aubert:2007bx}. Signal \B\ decays, \bln, are selected in 
the recoil of a semileptonic decay, \btodlnu, referred to as the ``tag'' \B.
The final states of the \taup\ decay in \btn\
are identical to those in Ref.~\cite{Aubert:2007bx}:
$\tautoenunu$, $\tautomununu$, $\tautopinu$, and $\tautopipiznu$. For the 
first time, we include \ben\ and \bmun\ in this search. 
In addition to using about 20\% more data than in Ref.~\cite{Aubert:2007bx}, 
we relax the constraints on the tag \B, improve the definition of the 
discriminating variables and use a combination of tag and 
signal \B\ variables in a multivariate discriminant that improves signal efficiency
and background rejection.

The tag $B$ is reconstructed in the set of semileptonic $B$ decay modes 
\btodlnu, through the full hadronic reconstruction of \Dz\ mesons and
identification of the lepton, $\ell^-$, as either \en\ or \mun. Other particles
($X$) resulting from a transition from a higher-mass charm state down to
the \Dz\ are not explicitly reconstructed and are not included in the tag \B\
kinematics. This strategy, and the reconstruction method (\Dz\ decay modes,
$\Dz\ell^-$ vertex requirements, etc.), are the same as in Ref. \cite{Aubert:2007bx}.
One difference in the present analysis is that we may assign up to one
photon (from $X$) back to the tag \B, based on its consistency with the
decay $\Dstarz \to (\piz,\gamma) \Dz$.

The efficiency for tag \B\ reconstruction ($\varepsilon_{\rm tag}$) is defined
as the rate at which events in the signal MC are found to contain at least 
one reconstructed tag \B\ and a single track recoiling against that tag. 
The efficiency for each signal mode is
given in Table \ref{tab:corrected_efficiencies}, including
corrections for systematic effects (described below). The efficiency 
is larger for \btn\ events due to high-multiplicity \taup\ decays
faking tag \B\ mesons.

We identify one of the following 
reconstructed particles recoiling against
the tag \B: $e^{+}$, $\mu^{+}$, $\pi^{+}$, or $\rho^{+}$.  The  
$e^{+}$ and $\mu^{+}$ can come from  $\btn$, with the
$\taup$ decaying leptonically, 
or directly from $\bmun$ or $\ben$.
The signal track must originate from the 
interaction point (IP), with a distance of closest approach to the IP
less than 2.5~\cm\ along the beam axis and less than 1.5~\cm\ 
transverse to the beam axis. We reject events
that contain more than one such IP track recoiling against
the tag \B. There may be additional tracks that do not come
from the IP. We reject events where the single IP track is identified as a kaon. 
We assign the single-track recoils to categories based on a hierarchical 
selection. An event is assigned to the $\mup$ category if 
the track passes muon identification or to the $\ep$ category if 
it passes electron identification; in the latter category, we recover up to 
one bremsstrahlung photon based on angular separation from the track
and add its four-momentum to the electron's. We assign the event to
the $\rho^{+}$ category if it fails 
lepton identification and can be paired with a \piz\ candidate.
The \piz\ candidates used in the $\rho^{+}$ reconstruction are defined 
as a pair of photons, each with laboratory energy $>50\mev$, with
invariant mass $m_{\piz}=[0.115,0.150]\gevcc$. Single-track
events that fail the selections above are assigned to the \pip\ 
category.

While the direction of neither \B\ meson can be known precisely,
four-momentum conservation constrains the tag \B\ momentum to lie on
a cone around the flight direction of the reconstructed $\Dz\ell^{-}$
system. The cosine of the opening angle between the \B\ meson and the 
$\Dz\ell^{-}$ system in the CM frame is given by
\begin{equation}
\label{eq:CosBY}
\cos\theta_{B,Y} = \frac{2 E_{B} E_{Y} - m_{B}^{2} - m_{Y}^{2}}{2|\vec{p}_{B}||\vec{p}_{Y}|},
\end{equation}
where $Y$ refers to the reconstructed tag \B\ final state,
($E_{Y}$, $\vec{p}_{Y}$) and
($E_{B}$, $\vec{p}_{B}$) are the 
four-momenta in the CM frame, and $m_{Y}$ and $m_{B}$ 
are the masses of the $Y$ system and tag $\Bu$ meson, respectively. 
$E_{B}$ and the magnitude of $\vec{p}_{B}$ are calculated 
from the beam energy: $E_{B} = E_{\rm{CM}}/2$ and 
$ | \vec{p}_{B} | = \sqrt{E_{B}^{2} - m_{B}^{2} }$.
Decays of the \B\ meson directly to $\Dz\ell^{-}\nu$ are
largely constrained to the physical region of this cosine,
while decays involving a higher-mass charm state will yield cosine
values below the physical region when the intermediate decay
particles (e.g. \piz\ or $\gamma$) are not explicitly reconstructed.

The signal \B\ momentum vector is equal in magnitude to $|\vec{p}_B|$ and 
is opposite to the tag \B\ direction, so that it lies on the cone of the 
tag \B\ momentum defined by Equation \ref{eq:CosBY}.
To estimate quantities in the signal \B\ rest frame, such as the momentum
of the signal \B\ daughter(s), we choose the signal \B\ boost vector on that cone
and we compute the quantity in the corresponding rest frame. We then use the value of that 
quantity averaged over all trial rest frames as an estimate of the true value. 
We denote the momentum of the signal particle(s)
determined by this method as $\pprimesignal$. 
This has the largest impact in the $\ben$ and $\bmun$ channels, where the 
lepton is mono-energetic in the signal \B\ rest frame. The improved 
resolution of the lepton momentum directly improves the 
separation of signal and background.
If an event has a reconstructed signal muon (electron) 
candidate and $\pprimesignal > 2.30\,(2.25) \gevc$, 
it is classified as a $\bmun$ ($\ben$) candidate; 
otherwise, it is classified as \btn, 
with $\tautomununu$ ($\tautoenunu$).

A critical discriminating variable is the extra energy (\eextra),
which is the total energy of charged and neutral particles that
cannot be directly associated with the reconstructed daughters 
of the tag \B\ or the signal \B. 
This variable was not examined (kept ``blind'')
until the analysis strategy was finalized.
We expect the signal to concentrate near zero $\eextra$; however, due to
collider-induced backgrounds,
detector noise, and unassigned tracks and neutrals from the tag and
signal \B\ mesons,
signal events can have non-zero \eextra. 
We require a minimum energy in the laboratory frame 
of 30\mev\ for any neutral cluster used in
\eextra. We improve our signal and background 
separation in this variable by using
an algorithm to assign up to one photon from the \eextra\ back to the tag \B.   
Candidate extra photons must
have a CM-frame energy less than 300 \mev,
consistent with having come from a \piz\ or 
$\gamma$ from the $\Dstarz \to \Dz$ transition. 
If, by adding a candidate photon back to the tag \B\ kinematics, 
the value of $\cos\theta_{B,Y}$  becomes  
closer to (but not greater than) 1.0, it is retained as a transition particle
candidate.  If more than one photon satisfies these 
conditions, the one which moves 
$\Delta M \equiv m_{\Dz\gamma} - m_{\Dz}$ closest to the nominal value of 
142 \mevcc \cite{pdg} is used.  This photon is excluded from $\eextra$. 
The tag \B\ kinematic quantities and \pprimesignal\ are recomputed, with
the photon added to the tag \B\ final state.

The background consists primarily of $B^{+}B^{-}$ events in which the tag
$B$ meson has been correctly reconstructed and the recoil contains
one reconstructed track and additional particles that are not 
reconstructed. 
Typically these events contain $K_{L}^{0}$ mesons and other particles
that are not detected and thus fake the multiple neutrinos in
signal events. Backgrounds from \B\ decays and continuum processes
have distinctive signatures in a number of discriminating quantities.
We group variables according to those which are computed from the whole event,
the tag \B, the signal \B, and other sources. Some variables, such as
those associated with the whole event, are useful for rejecting
continuum background, while others (such as those associated with 
the reconstructed \B\ mesons) are better at rejecting \B\ background.

The event-level variables are: 
the ratio of the second and zeroth Fox-Wolfram moments~\cite{Fox:1978vu};
the minimum invariant mass of any two charged tracks in the event;
the net charge of the event; $\cos\theta_{B,Y}$;
the invariant mass of the two leptons in the event ($m_{\ell\ell}$); and
the missing mass vs. cosine of the polar angle (laboratory frame)
of the missing three-momentum, where the sum defining the reconstructed 
four-momentum runs over all charged and neutral particles in the event. 
The tag \B\ variables are: 
the \Dz\ decay mode; the CM momenta of the
tag \B\ kaon and lepton;
particle identification quality of the tag \B\ charged kaon (where
applicable).
The signal \B\ variables are:
the quality of the particle identification of the signal muon,
for muon final states of the signal \B; 
the quality of
the kaon identification on the signal track (to reject
kaons misidentified as leptons or pions); for $\tautorho$,
the reconstructed mass of the $\rho^{+}$, and the CM momenta of the
$\rho^{+}$ daughters; and
for \btn, $\cos\theta^{\prime}_{\tau,Y}$ vs. \pprimesignal, where
$\cos\theta^{\prime}_{\tau,Y}$ is defined in the signal \B\ meson rest frame
using Equation \ref{eq:CosBY}, replacing \B\ meson quantities with those
of the $\tau$ ($E_{\tau}=m_{B}/2$ and $p_{\tau}=\sqrt{m_B^2-m_{\tau}^2}$) and
where $Y$ refers to the reconstructed $\tau$ 
final state (computed using the signal \B\ meson rest 
frame averaging procedure).
Other variables used are:
the separation between the tag \B\ meson decay vertex and the
point of closest approach to the IP by the
signal \B\ track; 
and the distribution of the cosine of the
angle between the signal \B\ CM momentum and the
tag \B\ thrust vs. the minimum invariant mass of
any three charged particles in the event~\cite{Aubert:2007bx}.

The shapes of these variables in MC simulation
are then used to define probability density functions
(PDFs) for signal ($P_s$) and background ($P_b$). We define
for each variable the ratio $P_s/[P_b + P_s]$.
We use the product of these ratios to construct a 
pair of likelihood ratios (LHRs) for 
each signal channel, one for rejecting \B\ backgrounds ($\lhrbb$) and the other
for rejecting continuum ($\lhrcont$) backgrounds. The LHR output
is bounded between 0 and 1, with signal accumulating toward 1 and 
background toward 0. 

We optimize selection criteria on $\eextra$, $\lhrbb$, and $\lhrcont$
for all modes. For the $\ben$ and $\bmun$ modes, we additionally optimize
the selection on $\pprimesignal$. For the  $\tautoenunu$ mode we
additionally optimize the selection on $m_{\ell\ell}$ (to reject
poorly modeled photon-conversion background). For the $\tau$ decay modes, 
we choose the figure-of-merit (FOM) to be $\Nsig/\sqrt{(\Nsig+\NBG)}$, since there is
still significant background left in these channels even after final selection
criteria are applied. For \bmun\ and \ben\ we use 
$\Nsig/(3/2+\sqrt{\NBG})$~\cite{Punzi:2003bu}
due to the low expected background.
We divide the MC simulation samples for signal and background into
thirds, two for optimization and one from which to compute
unbiased efficiencies and background predictions. This latter
sample has statistics roughly equivalent to the data.
Optimized selection criteria are
given in Table \ref{tab:optcuts}. The signal efficiency
($\varepsilon_{\rm sig}$)
is defined as the rate at which signal events containing
a reconstructed tag \B\ are also found to contain a signal \B\
candidate, and it includes the \taup\ branching fractions.
These efficiencies are given in 
Table \ref{tab:corrected_efficiencies}.

\begin{table}[htb]
\caption{Optimized signal selection criteria.}
\label{tab:optcuts} 
\footnotesize
\begin{center} 
\begin{tabular}{|l|c|c|c|c|c|} \hline
Mode            &  \lhrbb  & \lhrcont & $\eextra$ & $\pprimesignal$ & $m_{\ell\ell}$  \\ 
                &          &          & $(\gev)$  & $(\gevc)$       & $(\gevcc)$ \\ \hline
\multicolumn{6}{|c|}{\btn}\\ \hline
$\ep\nub\nu$	& $>0.77$  & $>0.25$  & $<0.20$            & -                    & $>0.29$ \\ 
$\mup\nub\nu$	& $>0.14$  & $>0.72$  & $<0.24$            & -                    & -  \\ 
$\rho^{+}\nu$	& $>0.97$  & $>0.95$  & $<0.24$            & -                    & -  \\ 
$\pip\nu$	& $>0.57$  & $>0.80$  & $<0.35$            & -                    & -  \\ \hline
\multicolumn{6}{|c|}{$\Bp \to (\mup,\ep) \nu$}\\ \hline
$\mup\nu$	& $>0.33$  & $>0.61$  & $<0.72$            & $[2.45,2.98]$          & -  \\ 
$\ep\nu$	& None     & $>0.01$  & $<0.57$            & $[2.52,3.02]$          & -  \\ \hline
\end{tabular}
\end{center}
\end{table}

We calibrate our background prediction using
sideband regions of \eextra\ where the signal 
contribution is negligible. We define the sidebands for \btn, \bmun, and \ben\  as 
$\eextra \ge 0.4 \gev$, $\ge 0.72 \gev$, and $\ge 0.6 \gev$, respectively.
We predict $\mathcal{N}_{\rm bg}^{\rm data}$, 
the number of background events in data in the
\eextra\ signal region (Table \ref{tab:BGpred_EEx}), 
by scaling the yield predicted by the MC simulation 
($N_{\rm bg}^{\rm MC}$) 
by the ratio of yields in data ($N_{\rm side}^{\rm data}$) and MC 
($N_{\rm side}^{\rm MC}$) in the sideband.
This method assumes that the shape of \eextra\ is well-described but does
not rely on the absolute prediction of the yield.
We validate this approach by defining sidebands in other variables 
($D^{0}$ mass, $\lhrcont$, $\lhrbb$, and $\pprimesignal$) and 
studying the data/MC agreement for the entire \eextra\ background
shape. We find the shape to be well described. 
We also studied the effect of varying the \eextra\ sideband definition
and obtained consistent background predictions.

\begin{table}
\caption{Background predictions from the $\eextra$ sideband, as described
in the text.}
\label{tab:BGpred_EEx}
\begin{center}
\begin{tabular}{|l|c|c|c|c|}\hline
Mode & $N_{\rm side}^{\rm MC}$ &	$N^{\rm data}_{\rm side}$               & $N^{\rm MC}_{\rm bg}$ & $\mathcal{N}^{\rm data}_{\rm bg}$ \\ \hline
$\tautoenunu$           &  333 $\pm$ 19  & 334  $\pm$ 18     & 81 $\pm$ 10            & 81 $\pm$ 12            \\
$\tautomununu$		&  1248 $\pm$ 36 & 1236 $\pm$ 35	&136 $\pm$ 12 	         & 135 $\pm$ 13	         \\
$\tautopinu$		&  6507 $\pm$ 88 & 7167 $\pm$ 85	&212 $\pm$ 19 	         & 234 $\pm$ 19	         \\
$\tautopipiznu$		&  1841 $\pm$ 48 & 1734 $\pm$ 42	&62 $\pm$ 9 	         & 59 $\pm$ 9	         \\ \hline
$\bmun$                 &  12 $\pm$ 5    & 14 $\pm$ 4	        & 12 $\pm$ 5             & 13 $\pm$ 8	         \\ \hline
$\ben$			&  26 $\pm$ 6    & 42 $\pm$ 6	        & 15 $\pm$ 5             & 24 $\pm$ 11	         \\
\hline
\end{tabular}
\end{center}
\end{table}

The branching fraction for any of the decay modes is
\begin{equation}\label{eqn:bf}
\BR (\bln) = \frac{N_{\rm obs} - \mathcal{N}_{\rm bg}^{\rm data}}{2 N_{\Bp\Bm} \eps_{\rm tag} \eps_{\rm sig}},
\end{equation} 
where $N_{\rm obs}$ is the total number of events observed in the signal region and
$N_{\Bp\Bm}$ is the total number of $\Y4S\to\BpBm$ decays in the data.   
The estimation of $N_{\Bp\Bm}$ has an uncertainty of 
1.1\%~\cite{Aubert:2002hc}.

Potential sources of significant systematic uncertainty in $\eps_{\rm tag}$ and
$\eps_{\rm sig}$ include the tag reconstruction rate,
the modeling of \eextra, and signal track and neutral reconstruction.
We use ``double-tagged'' events to study possible effects. 
Double-tagged events contain two fully reconstructed, independent, 
oppositely charged semileptonic tag \B\ decays. 
These double-tagged events are analogous to signal, 
in that every particle that can be
assigned to the original \B\ decays has been assigned.

We use the absolute yields
of tagged events to obtain a systematic uncertainty on $\eps_{\rm tag}$. 
We form a double ratio from the ratios of double-tagged to single-tagged
events in the data and MC simulation. Single-tagged events are defined
as events containing at least one semileptonic tag \B\ decay with no constraints
on the rest of the event. We improve the sample purity by requiring that
$D^{0} \to K^{-}\pi^{+}$ in at least one of the tags. 
We measure this double ratio to be $0.891 \pm 0.021$.
As a comparison, we perform the same
measurement replacing $D^{0} \to K^{-}\pi^{+}$ with  $D^{0} \to K^{-}\pi^{+}\pi^{-}\pi^{+}$
and find the double-ratio to be $0.954\pm0.011$. 
We use 0.891 as the nominal correction to $\varepsilon_{\rm tag}$ and treat 
the relative difference between the two methods (7.1\%) as the systematic uncertainty.

\begin{figure}[htb] 
\includegraphics[width =0.8\linewidth,keepaspectratio]{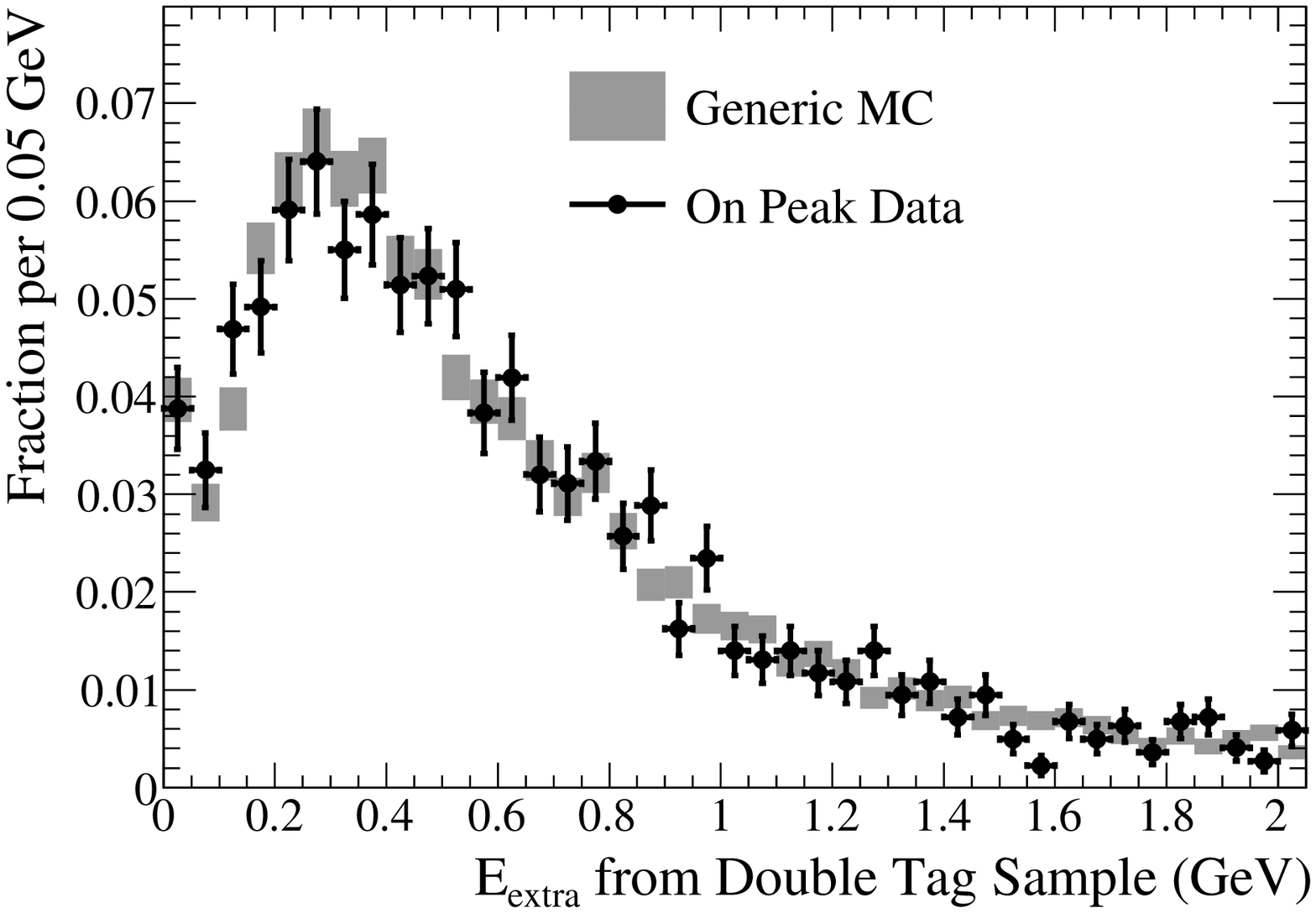}
\caption{\label{fig:DTEextra1456CutsUnit}Distribution of $\eextra$ in double-tagged events. The data (black
points) and MC simulated events (gray rectangles) are normalized to unit area. The  rectangles 
represent the MC simulation uncertainty.} 
\label{fig:DTEextra} 
\end{figure}

The \eextra\ distribution in double-tag events is expected to
contain contributions similar, though not identical, 
to those from signal events. We
validate \eextra\ using the double-tagged events described above,
additionally requiring that the second tag contains only
$\Dz\to K^- \pip$ and satisfies $\cos\theta_{B,Y}=[-1.1,1.1]$ 
to reject second tags with missing neutrals. The
resulting \eextra\ distribution is shown in 
Fig. \ref{fig:DTEextra1456CutsUnit}. 
It is well-described by the MC simulation.  We compare
the efficiency of selecting events in data and MC simulation 
for $\eextra \leq$ 0.4\gev\ and find that the
efficiency needs to be corrected  by $0.985 \pm 0.044$ 
to match the data. The uncertainty on this correction
is due to the statistical uncertainty on the data and MC simulation, 
and we treat it as a systematic uncertainty.

The remaining systematic uncertainties on $\eps_{\rm sig}$ come 
from tracking efficiency ($0.36\%$ per signal track),
 $\pi^{0}$ reconstruction for the $\tautopipiznu$ mode ($0.984 \pm 0.030$),
and particle identification. These are evaluated
using control samples of well-characterized particles.
The particle identification efficiency corrections and
systematic uncertainties are $0.953 \pm 0.003$ ($0.97 \pm 0.04$) for identified electrons in
the \btn\ (\ben) analysis and $0.92 \pm 0.05$ ($1.016 \pm 0.022$) for identified muons in
the \btn\ (\bmun) analysis.

\begin{table}[htb]
\caption{\label{tab:corrected_efficiencies}%
The corrected tag and signal efficiencies. The first uncertainty
is the MC statistical uncertainty, and the second is the 
systematic uncertainty from sources described in the text.
Branching fractions are included (e.g. $\taup\to\ep\nu\nub$).
The last column is the total systematic
uncertainty on each efficiency as a percent of its value.}
\renewcommand{\arraystretch}{1.3}
\begin{center}
\begin{tabular}{|c|c|c|}
\hline

Channel	&	Efficiency (\%)								&	Uncertainty (\%) \\ \hline\hline	
\multicolumn{3}{|c|}{Tag Efficiencies} \\ \hline
\btn	&	$(	1.514	\pm	0.003	\pm	0.107	)	$	&	7.1	\\	\hline
\bmun	&	$(	0.937	\pm	0.003	\pm	0.066	)	$	&	7.1	\\	\hline
\ben	&	$(	0.974	\pm	0.003	\pm	0.069	)	$	&	7.1	\\	\hline\hline
\multicolumn{3}{|c|}{Signal Efficiencies} \\ \hline
\tautoe	&	$(	1.58	\pm	0.04	\pm	0.07	)	$	&	4.5	\\	\hline
\tautomu	&	$(	1.45	\pm	0.03	\pm	0.11	)	$	&	7.4	\\	\hline
\tautopi	&	$(	2.44	\pm	0.05	\pm	0.11	)	$	&	4.5	\\	\hline
\tautopipiznu	&	$(	0.83	\pm	0.03	\pm	0.05	)	$	&	5.4	\\	\hline
\btn	&	$(	6.31	\pm	0.07	\pm	0.34	)	$	&	5.4	\\	\hline\hline
\bmun	&	$(	28.65	\pm	0.34	\pm	1.75	)	$	&	6.1	\\	\hline
\ben	&	$(	37.01	\pm	0.38	\pm	1.84	)	$	&	5.0	\\	\hline
\end{tabular}
\end{center}
\end{table}

\begin{figure}[htb]
\begin{center}
\hspace{-0.1in}
     \subfigure{\includegraphics[width=\twowidefig \linewidth]{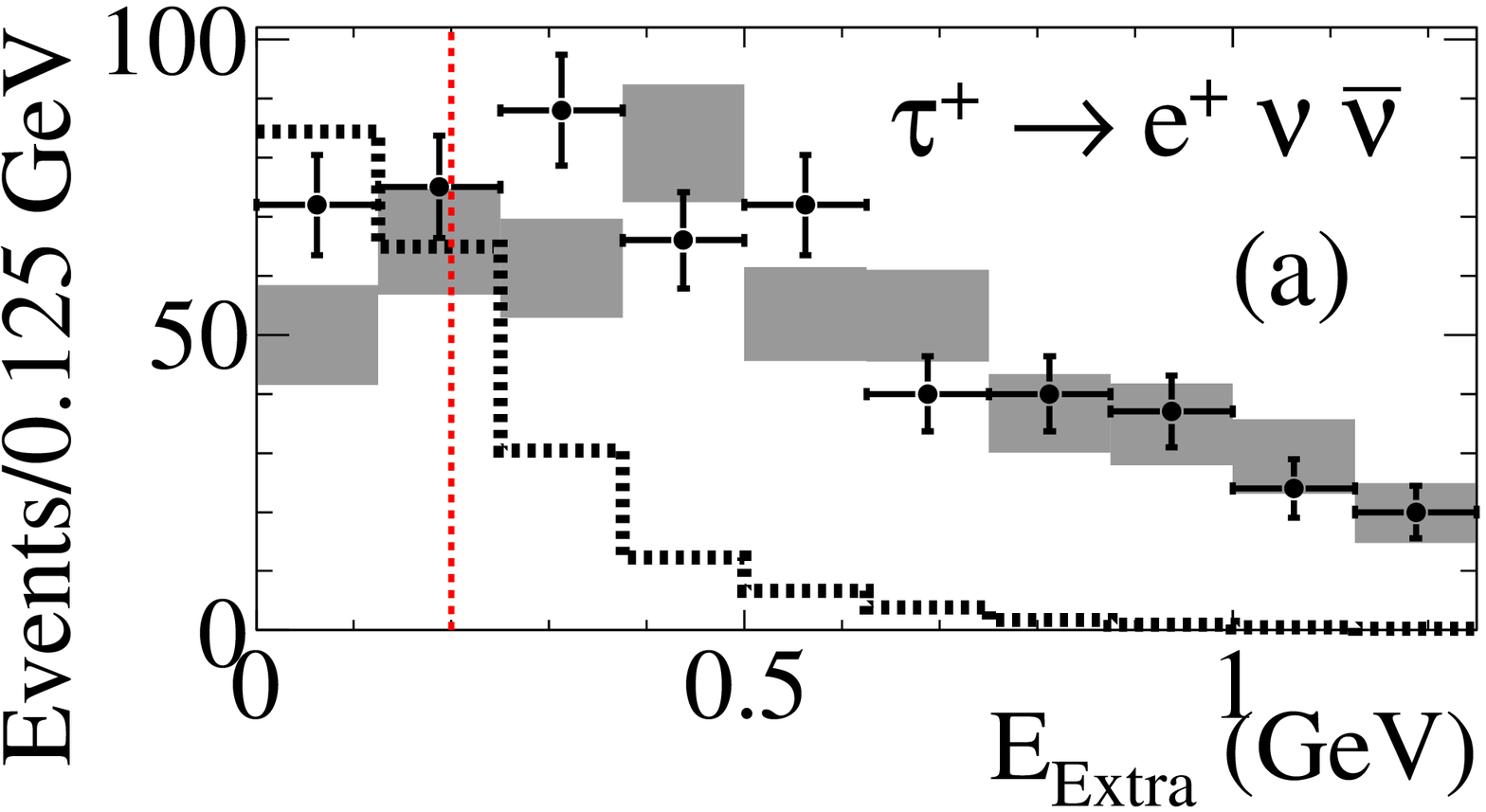}} 
     \hspace{.02in}
     \subfigure{\includegraphics[width=\twowidefig \linewidth]{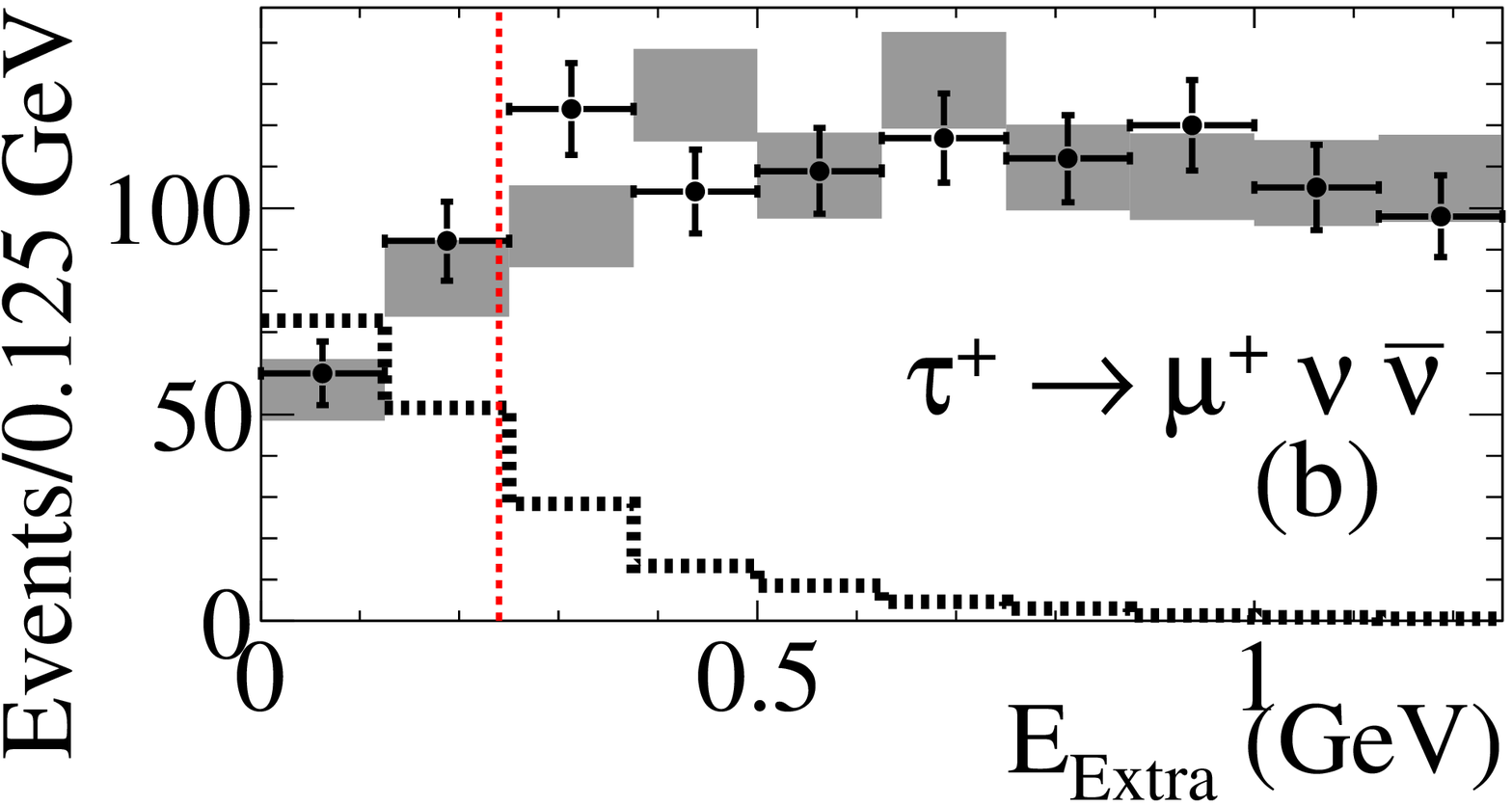}} 

     \subfigure{\includegraphics[width=\twowidefig \linewidth]{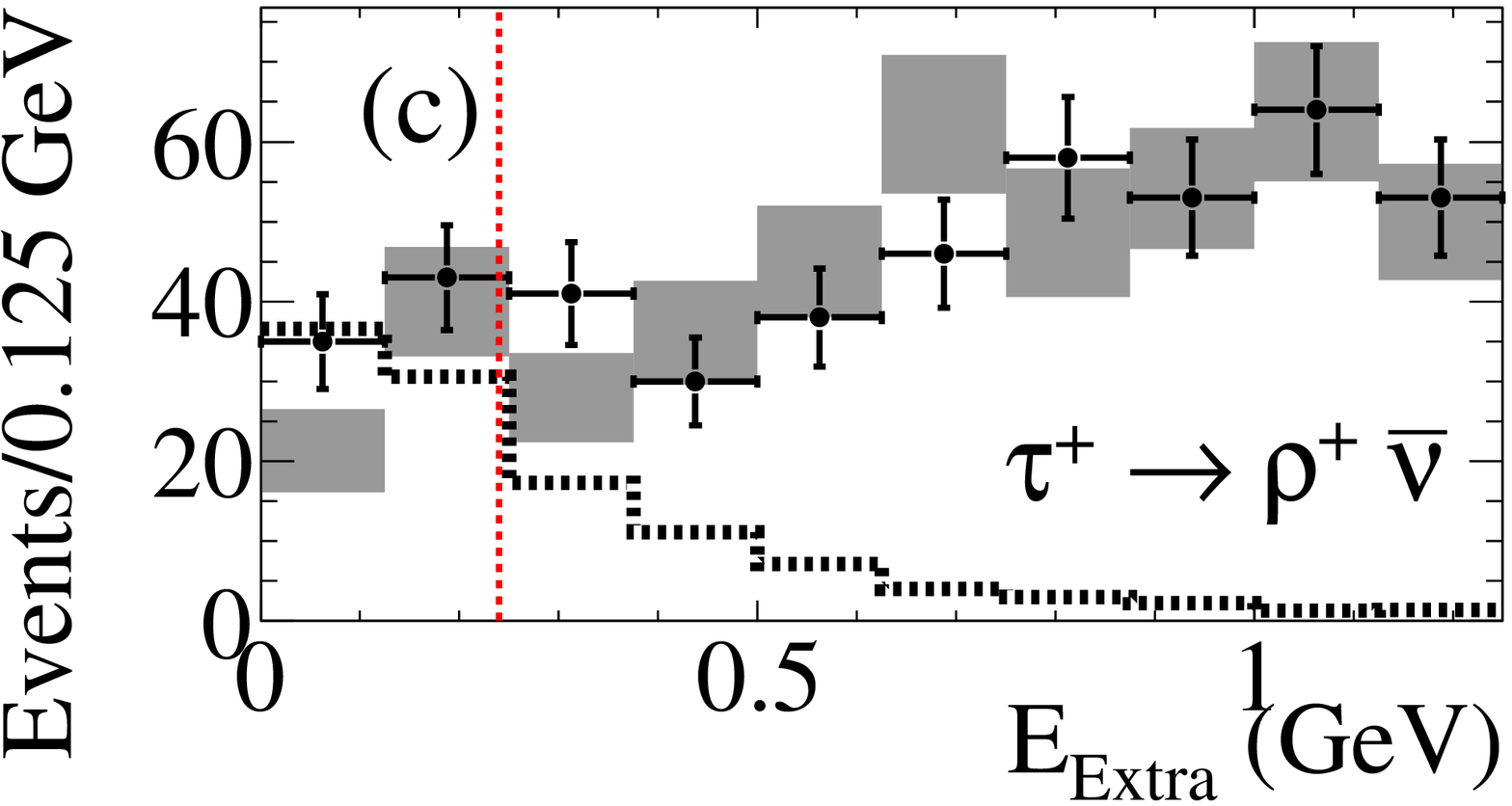}} 
     \hspace{.02in}
     \subfigure{\includegraphics[width=\twowidefig \linewidth]{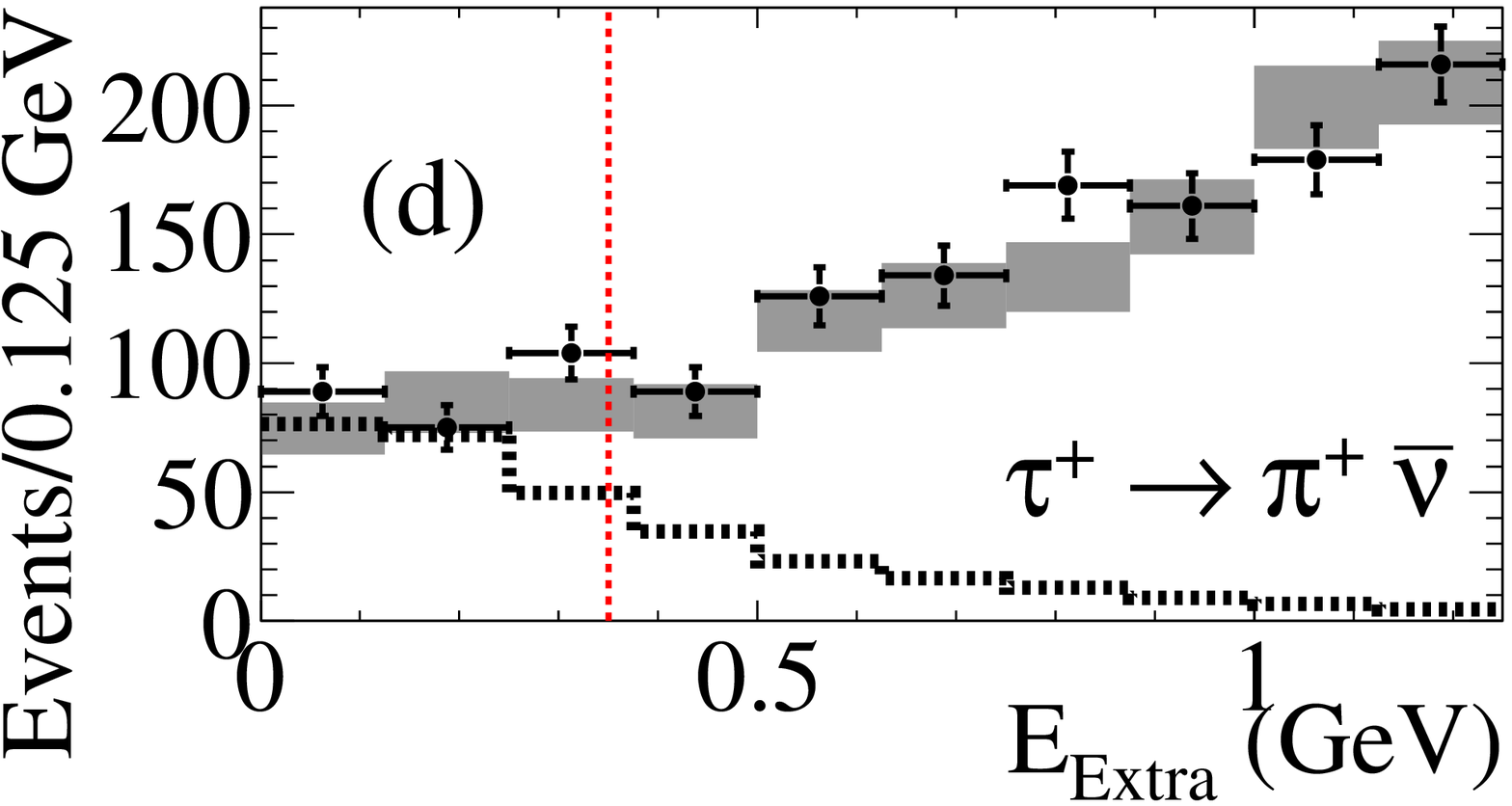}} 

     \subfigure{\includegraphics[width=\twowidefig \linewidth]{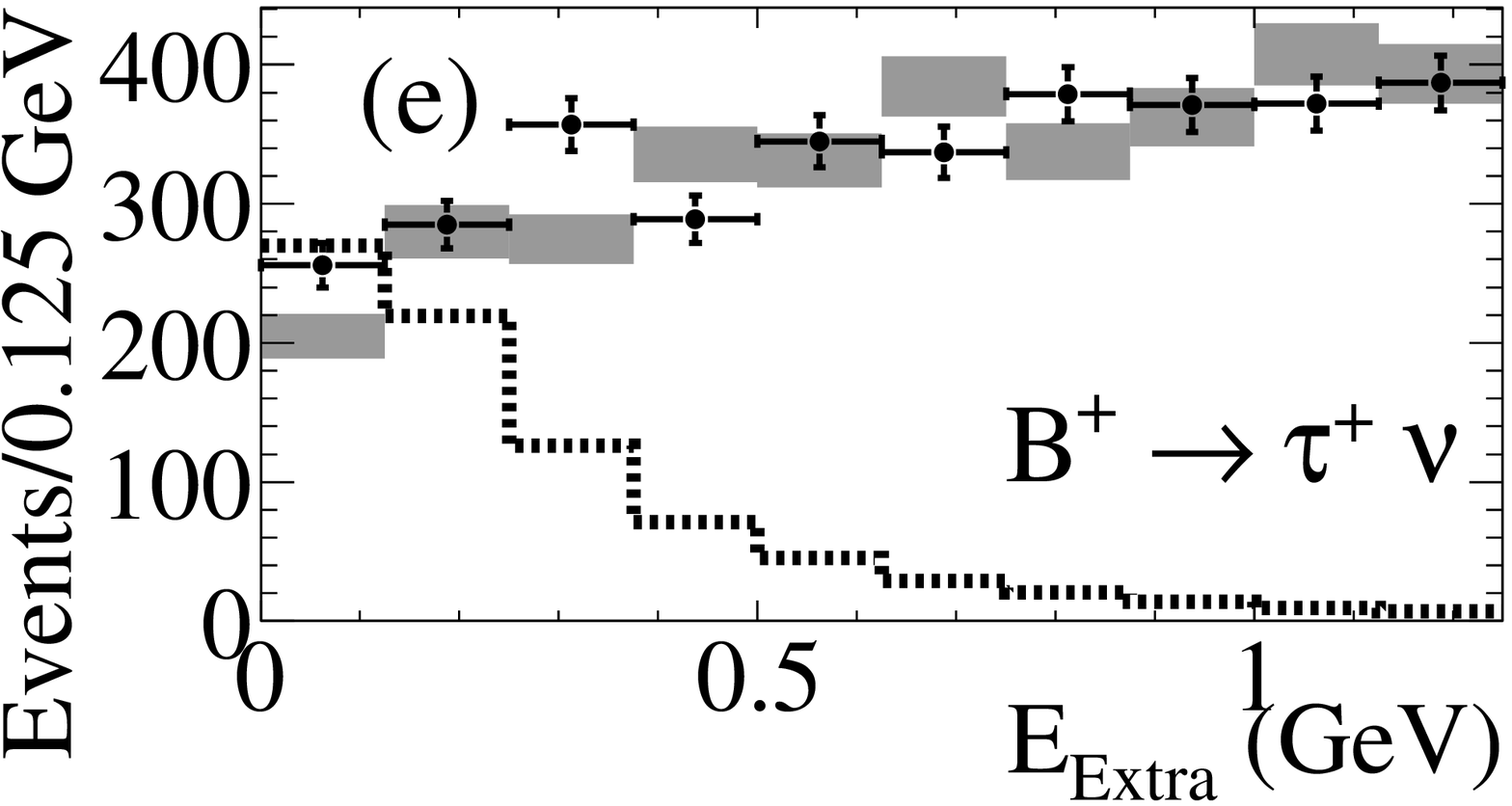}} 

     \subfigure{\includegraphics[width=\twowidefig \linewidth]{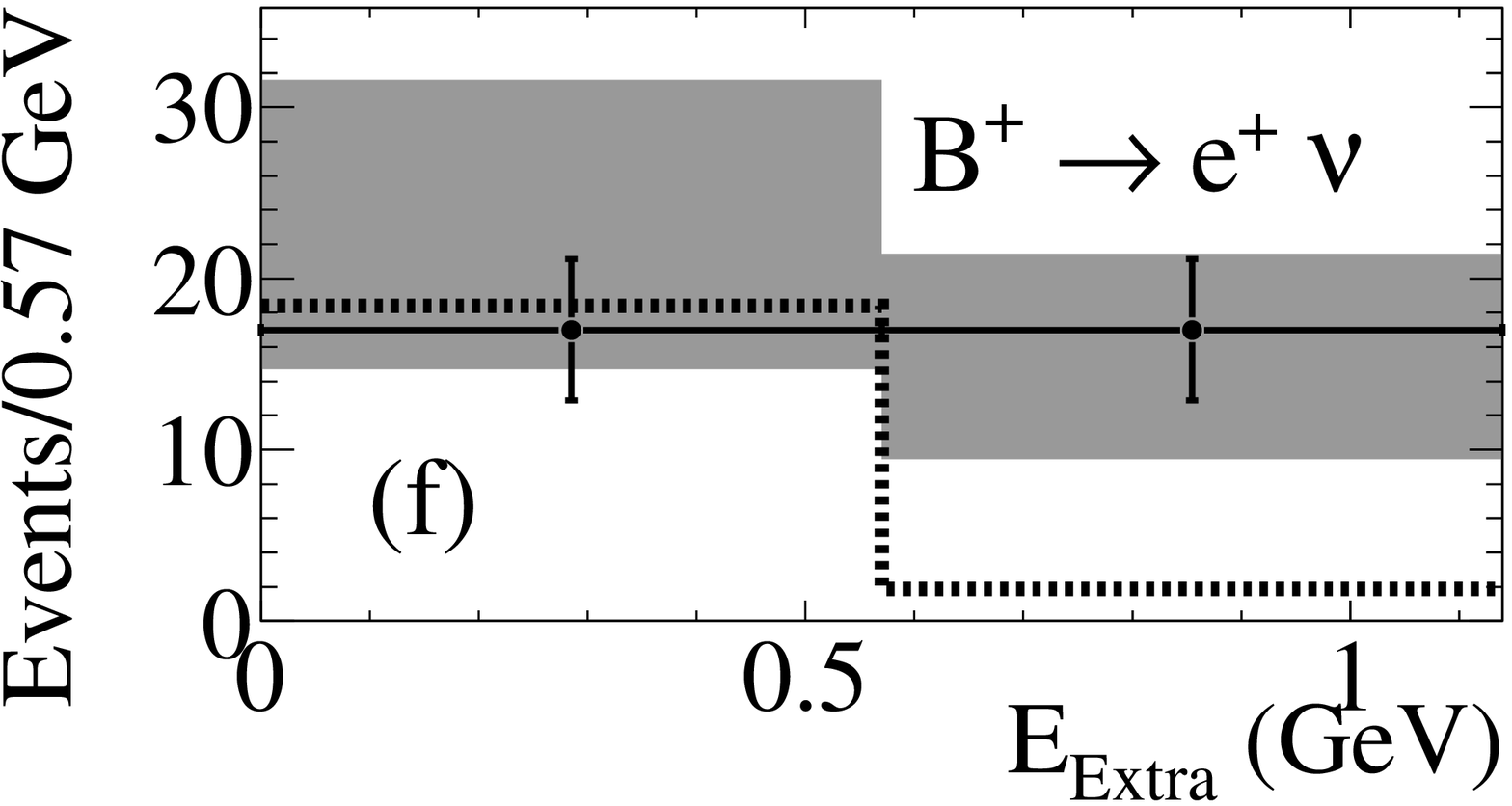}} 
     \hspace{.02in}
     \subfigure{\includegraphics[width=\twowidefig \linewidth]{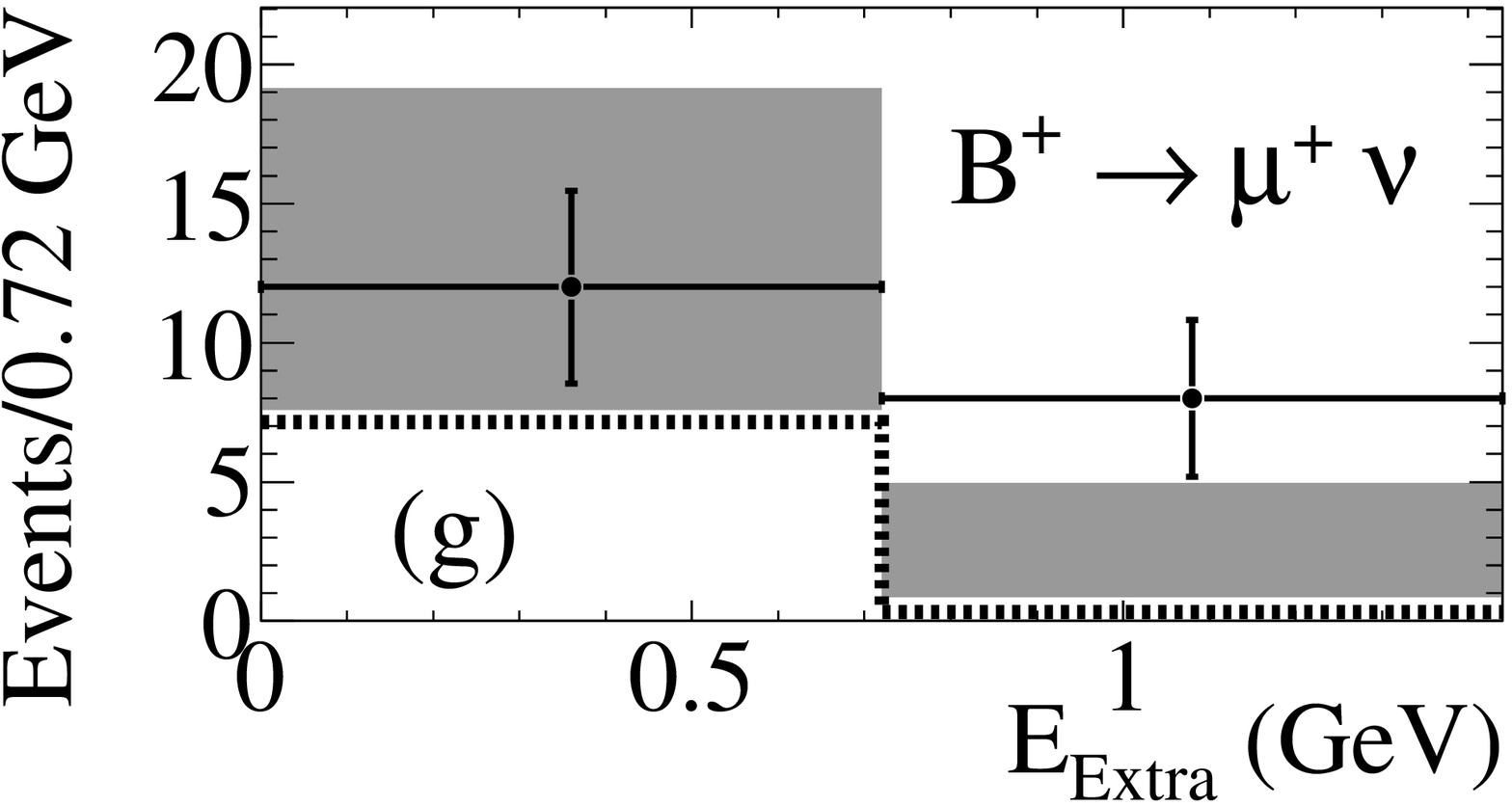}} 

\end{center}
\caption{\label{fig:unblinded_taunu}\eextra\ after all selection criteria have been 
applied for each final state. Shown are
data (black points), background MC simulation (gray shaded), and signal MC simulation
(dotted line) normalized to 10 times the expected branching fraction ($10^6$ times
for \ben). 
The background MC simulation is luminosity normalized and corrected for 
the data/MC ratio in the \eextra\ sideband; the rectangles represent the 
MC simulation statistical uncertainty. In (a-d), the vertical dashed line indicates 
the signal region boundary. In (f-g) the first bin is the signal region.} 
\end{figure}

The \eextra\ distributions for each channel are given in
Fig. \ref{fig:unblinded_taunu} and results 
given in Table~\ref{tab:Results}. We use the method of
Feldman and Cousins~\cite{FeldmanCousins} to interpret the yields
in each channel. When computing the level at which we
exclude the null hypothesis, we include systematic errors as a Gaussian
convolution with the nominal Poisson distribution. 
Our results in the \bmun\ and \ben\ channels are 
consistent with the background expectation and we obtain only
one-sided 90\% confidence intervals. 
For \btn, we
obtain a two-sided 68\% confidence interval and exclude the
null hypothesis at the level of $\btnsig$.
This result supersedes that of the previous work~\cite{Aubert:2007bx}.
The statistical consistency test of the results over the four \btn\ channels 
has a $\chi^{2}$ per degree-of-freedom 
of $2.02/3$, or a probability of 57\%, and is performed
using branching fractions computed with Equation \ref{eqn:bf}.
In the context of the SM we determine that 
$\fBSq = \fBSqresult$, where the uncertainty arises dominantly
from this measurement and $|V_{ub}|$.

\begin{table}
\caption{The expected background, observed events in data, and branching fraction results,
determined as described in the text.\label{tab:Results}}
\begin{center}
\begin{tabular}{|c|c|c|c|} \hline
Mode	& $\mathcal{N}^{data}_{bg}$  & $\Nobs$ & Branching \\
	&                            &         & fraction $(\times 10^{-4})$  \\ \hline\hline
$\tautoenunu$	&	81	$\pm$	12 &	121 & \btntautoerawresult	\\
$\tautomununu$	&	135	$\pm$	13 &	148 & \btntautomurawresult	\\
$\tautopipiznu$	&	59	$\pm$	9  &	71  & \btntautorhorawresult \\ 
$\tautopinu$	&	234	$\pm$	19 &	243 & \btntautopirawresult	\\ \hline
$\btn$		&	509	$\pm$	30 &	583 &  \btnrawresult	\\ \hline
$\bmun$	        &	13	$\pm$	8  &  12  & $< 0.11$ (90\% C.L.)\\ \hline
$\ben$		&	24	$\pm$	11 &	17  & $<  0.08$ (90\% C.L.)\\ \hline
\end{tabular}
\end{center}
\end{table}

We obtain a single \babar\ result for \btn\ by combining this result
with $\BR(\btn) = (1.8^{+1.0}_{-0.9}) \times 10^{-4}$, which is derived
from a statistically-independent sample using tag \B\ mesons decaying into 
fully hadronic final states~\cite{babar_hadronic_btn}.
We use a simple error-weighted average, since
the correlated systematics (mainly due to particle identification, charged particle
tracking, and \eextra) have a negligible impact on the combination.
We obtain $\BR(\btn) = \btnresultcomb$, which excludes zero at the $\combsig$ level.
Both this and the combined results are consistent with the SM prediction.

In conclusion, we have used the complete \babar\ data sample
to search for the purely leptonic \B\ meson 
decay $\Bu \to \ell^{+} \nu$
using a semileptonic \B\ decay tagging technique. 
We measure $\mathcal{B}(\btn) = \btnresult$ and exclude 
the null hypothesis at the level of $\btnsig$. We find 
results consistent with the background predictions for 
the decays \bmun\ and \ben. 

We are grateful for the excellent luminosity and machine conditions
provided by our \pep2\ colleagues, 
and for the substantial dedicated effort from
the computing organizations that support \babar.
The collaborating institutions wish to thank 
SLAC for its support and kind hospitality. 
This work is supported by
DOE
and NSF (USA),
NSERC (Canada),
CEA and
CNRS-IN2P3
(France),
BMBF and DFG
(Germany),
INFN (Italy),
FOM (The Netherlands),
NFR (Norway),
MES (Russia),
MEC (Spain), and
STFC (United Kingdom). 
Individuals have received support from the
Marie Curie EIF (European Union) and
the A.~P.~Sloan Foundation.

\bibliographystyle{apsrev}
\bibliography{paper}

\end{document}